\documentclass[aps,prb,twocolumn,showpacs]{revtex4}
\usepackage{amssymb}
\usepackage{graphicx}

\begin{document}

\title{Superconductivity and phase diagrams in 4d- and 5d-metal-doped iron arsenides SrFe$_{2-x}$M$_x$As$_2$ (M = Rh, Ir, Pd)}

\author{Fei Han, Xiyu Zhu, Peng Cheng, Gang Mu, Ying Jia, Lei Fang, Yonglei Wang, Huiqian Luo, Bin Zeng, Bing Shen, Lei Shan, Cong Ren and Hai-Hu Wen}\email{hhwen@aphy.iphy.ac.cn }

\affiliation{National Laboratory for Superconductivity, Institute of
Physics and Beijing National Laboratory for Condensed Matter
Physics, Chinese Academy of Sciences, P. O. Box 603, Beijing 100190,
China}

\begin{abstract}
By substituting the Fe with the 4d and 5d-transition metals Rh, Ir
and Pd in SrFe$_2$As$_2$, we have successfully synthesized a series
of superconductors SrFe$_{2-x}$M$_x$As$_2$ (M = Rh, Ir and Pd) and
explored the phase diagrams of them. The systematic evolution of the
lattice constants indicated that part of the Fe ions were
successfully replaced by the transition metals Rh, Ir and Pd. By
increasing the doping content of Rh, Ir and Pd, the
antiferromagnetic state of the parent phase is suppressed
progressively and superconductivity is induced. The general phase
diagrams were obtained and found to be similar to the case of doping
Co and Ni to the Fe sites. However, the detailed structure of the
phase diagram, in terms of how fast to suppress the
antiferromagnetic order and induce the superconductivity, varies
from one kind of doped element to another. Regarding the close
values of the maximum superconducting transition temperatures in
doping Co, Rh and Ir which locate actually in the same column in the
periodic table of elements but have very different masses, we argue
that the superconductivity is intimately related to the suppression
of the AF order, rather than the electron-phonon coupling.
\end{abstract} 
\maketitle

\section{Introduction}
The high temperature superconductivity found in
LaFeAsO$_{1-x}$F$_x$\cite{Hosono} was a surprising discovery since
the iron element in a compound in the most cases is a killer of
superconductivity due to its strong magnetic moment. In the
FeAs-based compounds, several different families have been found. In
the so-called 1111 phase with the ZrCuSiAs structure, the $T_c$ has
been quickly promoted to 56 K in thorium doped oxy-arsenide REFeAsO
(RE = rare earth elements)\cite{XuzaTh} and rare earth elements
doped fluoride-arsenide AeFeAsF (Ae = Ca,Sr)
compounds.\cite{xyzhu,cp} In the system of
(Ba,Sr)$_{1-x}$K$_x$Fe$_2$As$_2$ with the ThCr$_2$Si$_2$ structure
(denoted as 122 phase), the maximum T$_c$ at about 38 K was
discovered.\cite{BaKparent,Rotter,CWCh} This FeAs-122 phase provides
us a great opportunity to investigate the intrinsic physical
properties since large scale crystals can be grown.\cite{Canfield}
Furthermore, it was found that a substitution of Fe ions with Co can
also induce superconductivity with a maximum T$_c$ of about 24
K.\cite{Sefat,XuZA} Meanwhile, Ni substitution at Fe site in
BaFe$_2$As$_2$ has also been carried out with a T$_c$ of about 20.5
K.\cite{BaNiFeAs} This is very different from the cuprate
superconductors in which the superconductivity was always suppressed
when the Cu sites were substituted by other elements. Very recently,
superconductivity in Ru substituted BaFe$_{2-x}$Ru$_{x}$As$_{2}$ was
found.\cite{BaFe2-xRuxAs2} This indicates that, the
superconductivity can be induced by substituting the Fe with not
only the 3d-transition metals, such as Co and Ni, but also the
4d-transition metal, like Ru. Therefore, it is interesting to know
the results of substituting Fe ions with other 4d-transition metals
such as Rh and Pd which respectively locate below Co and Ni in the
periodic table of elements, as well as 5d-transition metal Ir
locating below Rh. In this paper, we report the successful
fabrication of the new superconductors SrFe$_{2-x}$M$_{x}$As$_{2}$
(M = Rh, Ir and Pd) by replacing the Fe with the 4d, 5d-transition
metals Rh, Ir and Pd. The maximum superconducting transition
temperatures were found at about 21.9 K in
SrFe$_{2-x}$Rh$_{x}$As$_{2}$, 24.2 K in
SrFe$_{2-x}$Ir$_{x}$As$_{2}$, and 8.7 K in
SrFe$_{2-x}$Pd$_{x}$As$_{2}$. X-ray diffraction (XRD) pattern, DC
magnetic susceptibility, resistivity, and upper critical field have
been determined on these 4d, 5d-transition metals doped
iron-arsenide superconductors. Based on these measurements, we get a
series of general phase diagrams corresponding to the different
doped transition metals.

\section{Sample preparation}
We synthesized the polycrystalline samples SrFe$_{2-x}$M$_x$As$_2$
(M = Rh, Ir and Pd) with a two-step solid state reaction
method.\cite{xyzhu2} Firstly, SrAs, FeAs and MAs (M = Rh, Ir and Pd)
were prepared by a chemical reaction involving Sr pieces, Fe powders
(purity 99.99\%), transition metal powders (purity 99.99\%) and As
grains (purity 99.99\%) together at 700 $^o$C for 20 hours. Then
these starting materials as well as Fe powders were mixed together
in the formula SrFe$_{2-x}$M$_x$As$_2$ (M = Rh, Ir and Pd), ground
and pressed into a pellet shape. All the weighing, mixing and
pressing procedures were performed in a glove box with a protective
argon atmosphere (both H$_2$O and O$_2$ are limited below 0.1 ppm).
The pellets were sealed in a silica tube under Ar gas atmosphere and
then heat treated at 900 $^o$C for 30 hours. Then they were cooled
down slowly to room temperature. A second sintering by repeating the
last step normally can improve the purity of the samples.

\section{Experimental data and discussion}
The X-ray diffraction (XRD) measurements of our samples were carried
out on a $Mac-Science$ MXP18A-HF equipment with a scanning range of
10$^\circ$ to 80$^\circ$ and a step of 0.01$^\circ$. The DC
magnetization measurements were done with a superconducting quantum
interference device (Quantum Design, SQUID, MPMS-7T). The resistance
data were collected using a four-probe technique on the Quantum
Design instrument physical property measurement system (Quantum
Design, PPMS-9T) with magnetic fields up to 9$\;$T. The electric
contacts were made using silver paste with the contacting resistance
below 0.05 $\Omega$ at room temperature. The data acquisition was
done using a DC mode of the PPMS, which measures the voltage under
an alternative DC current and the sample resistivity is obtained by
averaging these signals at each temperature. In this way the
contacting thermal power is naturally removed. The temperature
stabilization was better than 0.1$\%$ and the resolution of the
voltmeter was better than 10$\;$nV.

\subsection{X-ray diffraction}
\begin{figure}
\includegraphics[width=8cm]{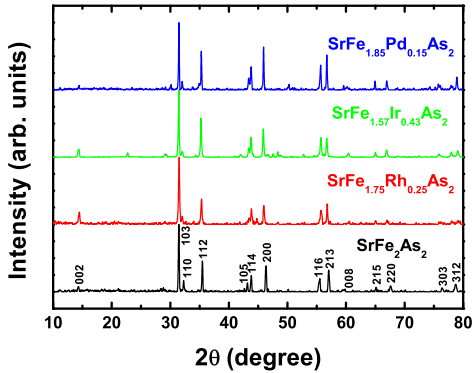}
\caption{(Color online) X-ray diffraction patterns of the samples
SrFe$_{2}$As$_{2}$, SrFe$_{1.75}$Rh$_{0.25}$As$_{2}$,
SrFe$_{1.57}$Ir$_{0.43}$As$_{2}$ and
SrFe$_{1.85}$Pd$_{0.15}$As$_{2}$. The latter three samples have the
optimized superconducting transition temperatures in their own phase
diagrams as shown below. Almost all main peaks can be indexed by a
tetragonal structure and the impurity phases are negligible. }
\label{fig1}
\end{figure}

\begin{figure}
\includegraphics[width=8cm]{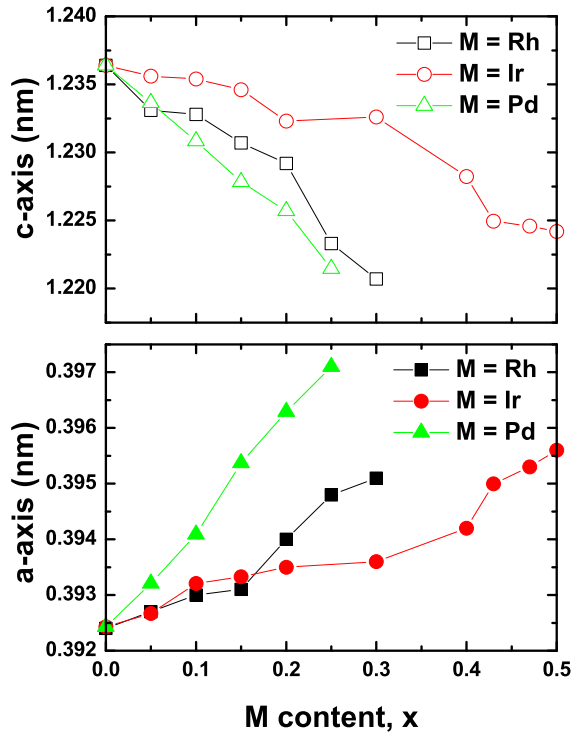}
\caption {(Color online) Doping dependence of the c-axis lattice
constant (top panel) and a-axis lattice constant (bottom panel). It
shows a common feature that the a-axis lattice constant expands,
while the c-axis one shrinks monotonically with Rh, Ir and Pd
substitution. This systematic evolution clearly indicates that the
Rh, Ir and Pd ions have been successfully substituted into the
Fe-sites. The $x$ here represents the nominal concentration of the
dopants.} \label{fig2}
\end{figure}

In Fig.~\ref{fig1} we present the x-ray diffraction patterns of the
samples SrFe$_{2}$As$_{2}$, SrFe$_{1.75}$Rh$_{0.25}$As$_{2}$,
SrFe$_{1.57}$Ir$_{0.43}$As$_{2}$ and
SrFe$_{1.85}$Pd$_{0.15}$As$_{2}$. The latter three samples have the
highest superconducting transition temperature in their own
families. All main peaks of the samples can be indexed to the
tetragonal structure very well and the impurity phases are
negligible. In order to have a comprehensive understanding to the
evolution induced by the doping process, we have measured the X-ray
diffraction patterns of almost all samples. By fitting the XRD data
to the structure with the software Powder-X,\cite{DongC} we get the
lattice constants of SrFe$_{2-x}$M$_{x}$As$_{2}$ (M = Rh, Ir and
Pd). The starting parameters for the fitting are taken from the
parent phase SrFe$_{2}$As$_{2}$\cite{ParentSrFe2As2} and the program
will finally find the best fitting parameters. In Fig.~\ref{fig2},
the a-axis and c-axis lattice parameters for the
SrFe$_{2-x}$M$_{x}$As$_{2}$ (M = Rh, Ir and Pd) samples were shown.
It is clear that by substituting the Rh, Ir, and Pd into the Fe
sites, the c-axis lattice constant shrinks, while the a-axis one
expands. This tendency is similar to the case of doping potassium to
the sites of Ba in Ba$_{1-x}$K$_x$Fe$_2$As$_2$, or substituting the
Fe with Ru in BaFe$_{2-x}$Ru$_{x}$As$_{2}$.\cite{BaFe2-xRuxAs2}
Normally a larger a-axis and smaller c-axis lattice constant would
mean that the bond angle of As-Fe-As is larger. A further refinement
of the structural data is underway. Concerning the very strong ZFC
diamagnetic signals as shown below, the XRD data here shows no doubt
that the bulk superconductivity arises from the
SrFe$_{2-x}$M$_{x}$As$_{2}$ (M = Rh, Ir and Pd) phase. We should
mention that the composition of Rh, Ir and Pd given here reflects
only the nominal value, as explained in next subsection.

\subsection{Scanning electron microscope analysis}

\begin{figure}
\includegraphics[width=8cm]{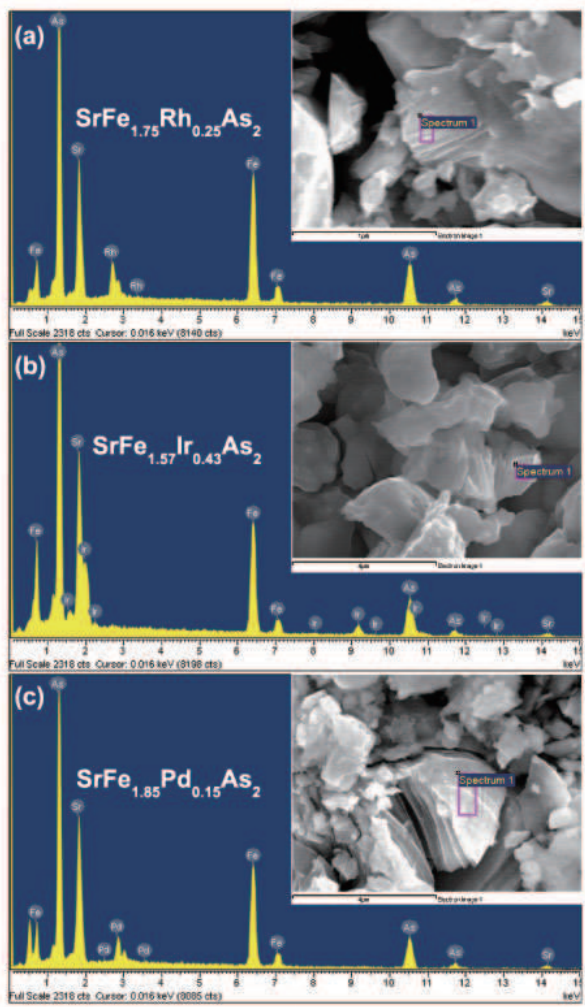}
\caption {(Color online) The Energy dispersive X-ray microanalysis
(EDX) spectrums of the the samples (a)
SrFe$_{1.75}$Rh$_{0.25}$As$_{2}$, (b)
SrFe$_{1.57}$Ir$_{0.43}$As$_{2}$ and (c)
SrFe$_{1.85}$Pd$_{0.15}$As$_{2}$. The spectrums are taken from the
main grains and show that the main elements of the grains are Sr,
Fe, M (M = Rh, Ir and Pd, respectively) and As. The insets show the
scanning electron microscopic pictures. The little rectangles mark
the positions where we took the EDX spectrums.} \label{fig3}
\end{figure}

Although the lattice constants change with the nominal doping
concentration systematically, it is still intriguing to check
whether the dopants (Rh, Ir and Pd here) are really doped into the
lattice, especially whether the true doping levels are close to the
nominal ones. Actually to obtain the chemical concentration of each
component in the sample is not a easy task. Here we adopt the simple
and fast way, using the Energy dispersive X-ray microanalysis (EDX)
spectra to do that. In the insets of Fig.3 (a)-(c) we present the
scanning electron microscope pictures of three typical samples with
the nominal formula SrFe$_{1.75}$Rh$_{0.25}$As$_{2}$,
SrFe$_{1.57}$Ir$_{0.43}$As$_{2}$ and
SrFe$_{1.85}$Pd$_{0.15}$As$_{2}$. As one can see, the grains in the
samples have irregular shapes and random sizes, but some have clear
layered structure. The EDX spectrum on the selected grains with
layered structures in the samples mentioned above are presented in
the main panel of Fig.3(a)-(c). In most cases, we can easily find
the expected component. The dopants (Rh, Ir and Pd) can be found in
the corresponding grains. Regarding to the relative concentrations
among the different components in the grains, the qualitative
consistency between the nominal concentration and the analyzed one
can still be followed. But the error bars of the analyzed
concentrations are large (at least $20\%$ varying from grain to
grain). The analyzed results obtained from three different grains
with nominal formulas SrFe$_{1.75}$Rh$_{0.25}$As$_{2}$,
SrFe$_{1.57}$Ir$_{0.43}$As$_{2}$ and
SrFe$_{1.85}$Pd$_{0.15}$As$_{2}$ are given in Table-I. We can see
that the general trend of doping effect is followed quite well. For
example, the Ir-doped sample has a maximum $T_c$ at the nominal
doping level of about 0.43, the Ir concentration found from this
typical grain is really quite high. While the Pd-doped one shows an
optimized superconductivity at the doping level of 0.15, the
analyzed value is relatively lower. Since the EDX results give quite
large uncertainty about the concentration which also scatters a lot
from grain to grain, it is not meaningful to adopt the analyzed
values. Therefore in this paper we use the nominal composition,
instead of the analyzed one to present our data and discussion.

\begin{table}[!h]
\tabcolsep 0pt \caption{Weight and atomic ratio of the elements for
the samples SrFe$_{1.75}$Rh$_{0.25}$As$_{2}$,
SrFe$_{1.57}$Ir$_{0.43}$As$_{2}$ and
SrFe$_{1.85}$Pd$_{0.15}$As$_{2}$.} \vspace*{-12pt}
\begin{center}
\def\temptablewidth{0.5\textwidth}
{\rule{\temptablewidth}{1pt}}
\begin{tabular*}{\temptablewidth}{@{\extracolsep{\fill}}ccccccc}
Nominal &Element &Weight$\%$ &Atomic$\%$\\
\hline
       SrFe$_{1.75}$Rh$_{0.25}$As$_{2}$ &Sr &22.74 &18.21\\
       &Fe &34.75 &43.66\\
       &Rh &6.63 &4.52\\
       &As &35.87 &33.60\\
\hline
       SrFe$_{1.57}$Ir$_{0.43}$As$_{2}$ &Sr &23.54 &20.95\\
       &Fe &24.92 &34.80\\
       &Ir &14.80 &6.01\\
       &As &36.74 &38.24\\
\hline
       SrFe$_{1.85}$Pd$_{0.15}$As$_{2}$ &Sr &26.31 &21.40\\
       &Fe &31.21 &39.82\\
       &Pd &5.75 &3.85\\
       &As &36.72 &34.93\\
       \end{tabular*}
       {\rule{\temptablewidth}{1pt}}
       \end{center}
       \end{table}

\subsection{DC magnetization}
\begin{figure}
\includegraphics[width=8cm]{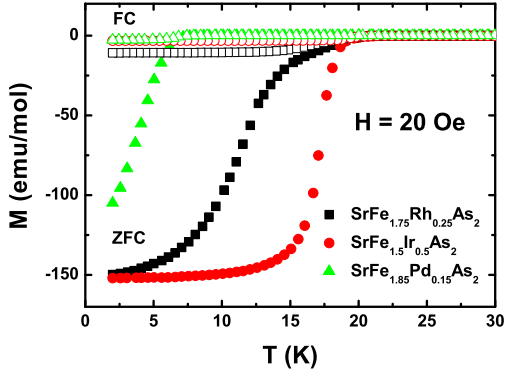}
\caption{(Color online) Temperature dependence of DC magnetization
for the samples SrFe$_{1.75}$Rh$_{0.25}$As$_{2}$,
SrFe$_{1.5}$Ir$_{0.5}$As$_{2}$ and SrFe$_{1.85}$Pd$_{0.15}$As$_{2}$.
The measurement was done under a magnetic field of 20 Oe with
zero-field-cooled and field-cooled modes. Strong diamagnetic signals
were observed here.} \label{fig4}
\end{figure}

In Fig.~\ref{fig4} we present the temperature dependence of DC
magnetization for the samples SrFe$_{1.75}$Rh$_{0.25}$As$_{2}$,
SrFe$_{1.5}$Ir$_{0.5}$As$_{2}$ and SrFe$_{1.85}$Pd$_{0.15}$As$_{2}$.
The measurement was carried out under a magnetic field of 20 Oe in
zero-field-cooled and field-cooled processes. Clear diamagnetic
signals appear below 21$\;$K for SrFe$_{1.75}$Rh$_{0.25}$As$_{2}$,
21.6 $\;$K for SrFe$_{1.5}$Ir$_{0.5}$As$_{2}$, and 8.2$\;$K for
SrFe$_{1.85}$Pd$_{0.15}$As$_{2}$, which correspond to the middle
transition temperatures of the resistivity data. The ZFC diamagnetic
signals are very strong in the low temperature regime. Although the
vortex pinning effect as well as the connectivity between the grains
give some influence on the diamagnetization signal, the strong
diamagnetization value here certainly signals a rather large volume
of superconductivity. However, we should point out that due to the
uncertainty in counting the issues mentioned above and the
demagnetization factor, it is difficult to calculate the precise
volume of superconductivity either from the ZFC or the FC
magnetization signal. For example, in the ZFC mode, if the
superconducting connectivity is good enough at the surface of a
superconductor, the ZFC signal may show a full screening effect, but
the inside may be non-superconductive. In the FC mode, the vortex
pinning can strongly influence the signal. For an uniform sample,
normally the stronger vortex pinning will lead to a smaller
diamagnetization signal. Regarding the polycrystalline feature of
our sample (without ideal superconducting connectivity at the
surface), the large diamagnetization signal measured here may only
point to a large superconducting volume.

\subsection{Resistivity and phase diagrams}
\subsubsection{SrFe$_{2-x}$Rh$_{x}$As$_{2}$ system}
\begin{figure}
\includegraphics[width=8cm]{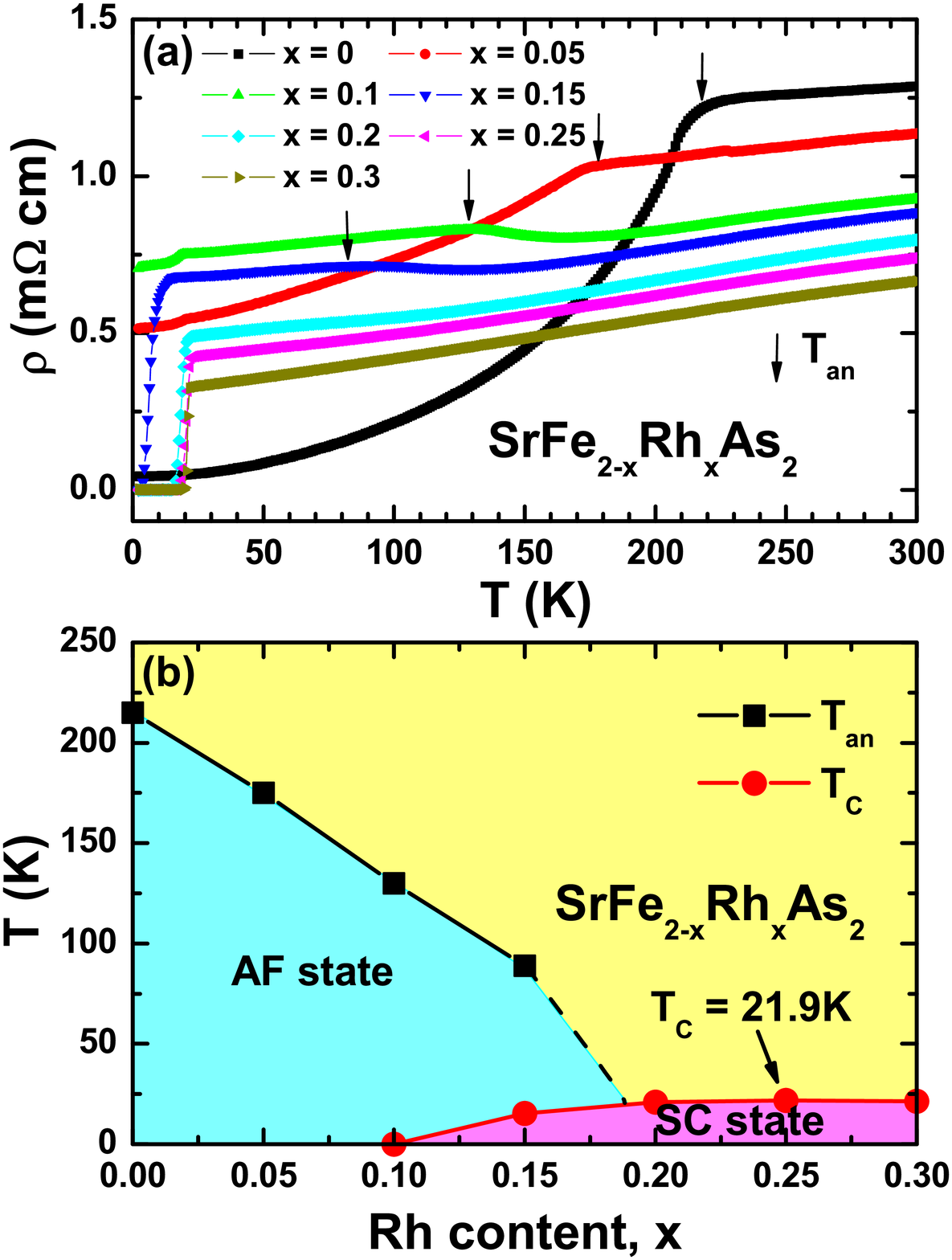}
\caption {(Color online) (a) Temperature dependence of resistivity
for samples SrFe$_{2-x}$Rh$_{x}$As$_{2}$ with x ranging from 0 to
0.3. The resistivity anomaly is indicated by the arrow for each
doping, which is determined as the onset of a kink in
resistivity-temperature curve. The tiny drop of resistivity at about
20 K for the sample x=0.10 may be induced by a small amount of
superconducting phase, suggesting slight inhomogeneity in the
sample. (b) Phase diagram of the superconductor
SrFe$_{2-x}$Rh$_x$As$_2$ with the Rh content x from 0 to 0.3. The
superconductivity starts to appear at x = 0.15, reaching a maximum
T$_c$ of 21.9 K at about x = 0.25. The dashed line provides a guide
to the eyes for the possible AF order/strctural transitions near the
optimal doping level.} \label{fig5}
\end{figure}

In Fig.~\ref{fig5}(a), we present the temperature dependence of
resistivity for samples SrFe$_{2-x}$Rh$_{x}$As$_{2}$. The parent
phase exhibits a sharp drop of resistivity (resistivity anomaly) at
about 215 K, which associates with the formation of the AF order. As
we can see, with more Rh doped into the
SrFe$_{2-x}$Rh$_{x}$As$_{2}$, the temperature of this anomaly was
suppressed (see, for example the sample x = 0.05). When x increases
to 0.15, superconductivity appears, while the anomaly still exists.
But here the resistivity anomaly shows up as uprising, instead of a
dropping down. This is slightly different from the case of Co
doping, where a very small amount of Co doping will convert this
sharp drop to an uprising. This difference may be induced by the two
effects which give opposite contributions to the resistivity in the
system: the decrease of the scattering rate as well as the charge
carrier densities. In the sample of x = 0.2, the resistivity anomaly
disappeared completely. Interestingly, the normal state resistivity
of the superconducting sample shows a roughly linear behavior
staring just above T$_c$ all the way up to 300 K. This is difficult
to be understood with the picture of phonon and impurity scattering.
It is certainly illusive to know whether this reflects an intrinsic
feature of a novel electron scattering. With x = 0.25, the maximal
T$_c$ at 21.9 K was found. The maximal transition temperature
appears at a higher doping level here (x = 0.25) compared with the
case of doping Co (x = 0.10-0.16). The underlying reason is unknown
yet. However it is interesting to mention that in the Ir-doped case
below, the maximal T$_c$ appears at about x = 0.43. It is yet to be
understood whether this is due to the evolution from doping with 3d
(Co), 4d (Rh) and 5d (Ir) transition metals so the superconductivity
comes later through 3d to 5d. We must mention that the absolute
value of resistivity derived from our polycrystalline samples here
may suffer a change from that of single crystals due to the grain
boundary scattering and the porosity. This happens quite often in
polycrystalline samples in which a larger resistivity was found when
compared with the single crystal sample. We also measured the
density of our three typical samples
SrFe$_{1.75}$Rh$_{0.25}$As$_{2}$, SrFe$_{1.5}$Ir$_{0.5}$As$_{2}$ and
SrFe$_{1.85}$Pd$_{0.15}$As$_{2}$. For these samples, the ideal
density calculated using the lattice constants determined in this
work is 6.288 g/cm$^3$, 7.088 g/cm$^3$ and 6.175 g/cm$^3$,
respectively, while the true density is 5.177 g/cm$^3$, 4.447
g/cm$^3$ and 4.764g/cm$^3$, respectively. Clearly the porosity
volume ratio can be as high as 20-30 \% in some samples. Therefore
the resistivity determined here, and perhaps also in general in all
other polycrystalline samples, can only tell us the qualitative
characteristics. Cations must be taken when using them to estimate
the intrinsic properties.

To build up the phase diagram for the three different dopants, we
determined the superconducting transition temperature T$_c$ value by
a standard method, i.e., using the crossing point of the normal
state background and the extrapolation of the transition part with
the most steep slope. Meanwhile the T$_{an}$ value was determined as
the onset point of the kink in the resistivity curve in the normal
state, which corresponds to the antiferromagnetic order. Based on
the data, we can get an electronic phase diagram for
SrFe$_{2-x}$Rh$_{x}$As$_{2}$ within the range of x = 0 to 0.3, which
is shown in Fig.~\ref{fig5}(b). Just like other samples in the
FeAs-122 family, with increasing Rh doping, the temperature of the
resistivity anomaly is driven down, and the superconducting state
emerges at x = 0.15, reaching a maximum T$_c$ of 21.9 K at x = 0.25.
The superconducting state even appears at the doping level of 0.3.
From the diamagnetization measurements, we found that this sample
has a much smaller superconducting volume compared with that of x =
0.25. As one can see, there exists a region in which the
antiferromagnetic and superconductivity coexists in the underdoped
side. This general phase diagram looks very similar to that of Co
doping.\cite{Canfield2,Fisher} Since Rh locates just below Co in the
periodic table of elements, we would conclude that the
superconductivity induced by Rh doping shares the similarity as that
of Co doping.

\subsubsection{SrFe$_{2-x}$Ir$_{x}$As$_{2}$ system}
\begin{figure}
\includegraphics[width=8cm]{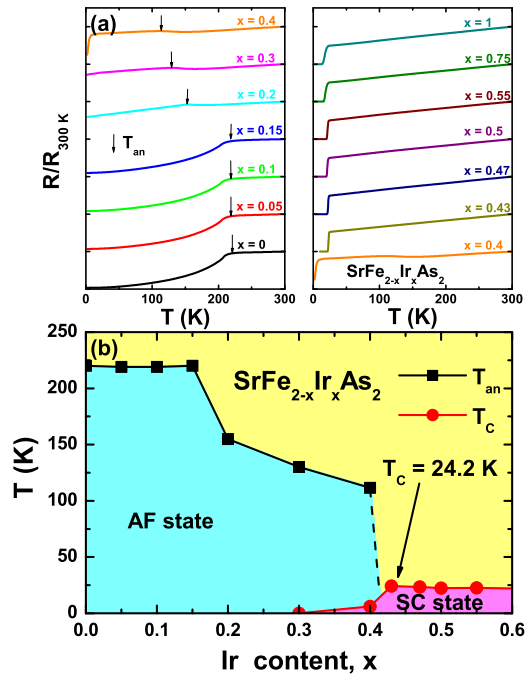}
\caption {(Color online) (a) Temperature dependence of resistivity
and (b) phase diagram for compounds SrFe$_{2-x}$Ir$_{x}$As$_{2}$
with the Ir content x from 0 to 1. The antiferromagnetic order of
the parent phase begins to be suppressed at x = 0.2. The
superconductivity starts to appear at x = 0.4, and reaches a maximum
$T_c$ of 24.2 K rapidly at about x = 0.43.} \label{fig6}
\end{figure}

Fig.~\ref{fig6}(a) shows the temperature dependence of resistivity
for samples SrFe$_{2-x}$Ir$_{x}$As$_{2}$  with x = 0 to 1,
respectively. It is interesting that the resistivity anomaly is not
suppressed while the doping level is increased from 0 to 0.15. In
this region, the varied a-axis and c-axis lattice indicate that the
Ir have been successfully doped into the Fe sites. When the doping
level gets higher (x $\geq$ 0.2), the temperature of the resistivity
anomaly T$_{an}$ begins to drop down, and the superconductivity
appears at the doping level of x = 0.4. In our superconducting
samples (x $\geq$ 0.43), the resistivity anomaly disappeared
completely. The sample with nominal composition x = 0.43 offers a
maximum superconducting transition temperature at about 24.2 K which
is determined in the same way as the Rh-doped case. The transition
width determined here with the criterion of 10-90 $\%$ $\rho_n$ is
about 1.7 K. With higher doping (x $\geq$ 0.47) the transition
temperature declines slightly. From the XRD data, we find that the
samples with higher doping levels (x $\geq$ 0.47) contain much more
impurities, therefore we are not sure whether this slight drop of
superconducting transition temperature is due to the chemical phase
separation or it is due to the systematic evolution of T$_c$ vs.
doping level. The normal state resistivity of the superconducting
samples (x $\geq$ 0.43) show a roughly linear behavior near the
optimized doping point, just like the Rh-doped case. Since the
sample with x = 0.43 shows already a reliable quality, we would
believe that this linear temperature dependence of resistivity is
intrinsic and may posses itself of great importance. More data are
desired to clarify this interesting feature in the normal state.

Both T$_{an}$ and T$_{c}$ were determined for each sample of
SrFe$_{2-x}$Ir$_{x}$As$_{2}$. Based on the data collected, we obtain
a general phase diagram, as shown in Fig.~\ref{fig6}(b). With
increasing the doping level, the T$_{an}$ is not driven down
immediately up to x = 0.2. With higher doping(x $\geq$ 0.2), the
antiferromagnetic order of the parent phase is suppressed, and there
exists a region in which the antiferromagnetic order and
superconductivity coexist in the underdoped side. When the doping
level reaches 0.43, the T$_c$ value is driven up to 24.2 K rapidly.
The superconducting state even appears in a wide overdoped region
from 0.47 to 1.0. Since some extra peaks from the impurity phase
appears for the sample with high doping, this clearly suggests that
there is a solubility limit of Ir doping. Therefore the phase
diagram was drawn only up to a nominal concentration of 0.60. The
general phase diagram looks similar to that of Co and Rh doping
since Ir locates just below Co and Rh in the periodic table of
elements. However, there are also several differences here compared
with that of Co and Rh doping. First the suppression to the AF order
is much weaker and it lasts to a quite high doping. The
superconductivity emerges suddenly at about 0.4 and reaches the
maximum T$_c$ at x = 0.43. Furthermore the T$_{an}$(x) curve is not
smooth in the underdoped region. Since here we just take this
anomaly from the resistivity, it may correspond to different
transitions in different doping regions. For example, in the low
doping region ($0\leq x \leq 0.20$), it may associate with the
AF/structural transition, while in the high doping region ($0.20\leq
x \leq 0.40$) this anomaly may correspond only to the structural
transition. Thus temperature dependent structural data are needed to
carry out the exact meaning of the T$_{an}$(x).

\subsubsection{SrFe$_{2-x}$Pd$_{x}$As$_{2}$ system}
\begin{figure}
\includegraphics[width=8cm]{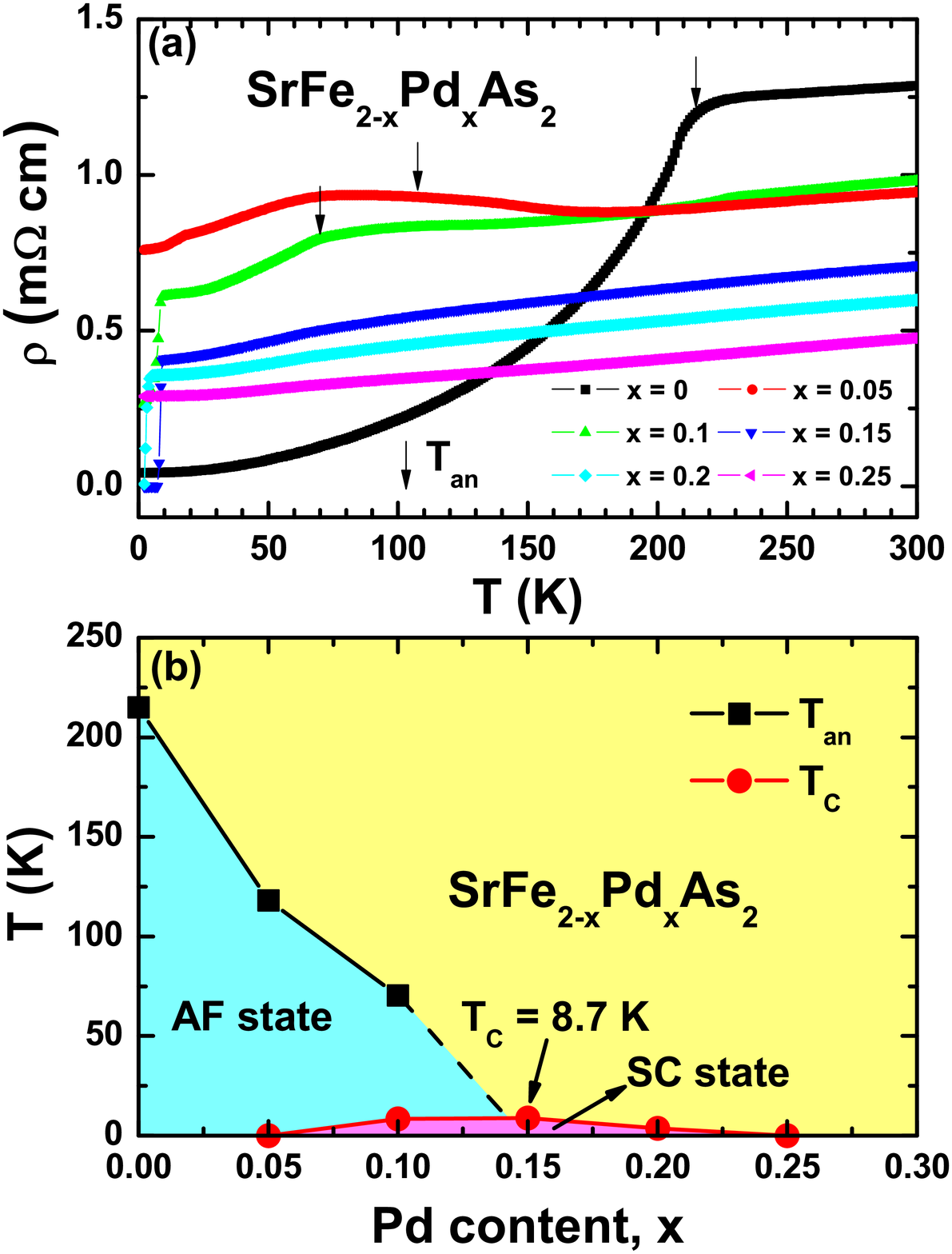}
\caption {(Color online) (a) Temperature dependence of resistivity
and (b) phase diagram for compounds SrFe$_{2-x}$Pd$_{x}$As$_{2}$
with the Pd content x ranging from 0 to 0.25. The superconductivity
starts to appear at x = 0.1, reaching a maximum T$_c$ of 8.7 K at x
= 0.15.} \label{fig7}
\end{figure}

In Fig.~\ref{fig7}(a) we present the temperature dependence of
resistivity for samples SrFe$_{2-x}$Pd$_{x}$As$_{2}$ with x = 0,
0.05, 0.1, 0.15, 0.20 and 0.25 respectively. By doping Pd to the Fe
sites, the resistivity-drop was converted to an uprising. This
occurs also in the Co, Rh and Ir-doped samples. We found that the
superconductivity appears in the sample with nominal composition of
x = 0.1. In the sample of x = 0.15, the resistivity anomaly
disappeared completely. It is found that the optimal superconducting
transition temperature is only about 8.7 K at a doping of x = 0.15.
The transition width determined here with the criterion of 10-90
$\%$ $\rho_n$ is about 1.2 K. With higher doping level (x = 0.2) the
transition temperature declines slightly. The superconductivity
again disappeared when the doping content x is over 0.25.

In Fig.~\ref{fig7}(b), a phase diagram of
SrFe$_{2-x}$Pd$_{x}$As$_{2}$ within the range of x from 0 to 0.25
was given. Just like the Rh and Ir doped samples, with increasing Pd
doping, the temperature of the anomaly is driven down, and the
superconducting state emerges at x = 0.1, reaching a maximum T$_c$
of 8.7 K at x = 0.15. The superconducting state disappeared at x =
0.25. As we can see, there exists an
antiferromagnetic-and-superconductivity-coexisting region in the
underdoped region. This is just like the Co and Rh doped cases, but
different from the Ir doped case. This general phase diagram looks
also similar to that of Ni doping.\cite{XuZA} Since Pd locates just
below Ni in the periodic table of elements, we would conclude that
the superconductivity induced by Pd doping shares the same mechanism
as that of Ni doping.

The maximum T$_c$ by doping Pd is only about 8.7 K while that of
other transition metal doped sample is much higher. It is still
unclear why the superconducting transition temperature varies in
doping different elements. In addition, in most cases, substituting
transition metal elements to the Fe sites in the 1111 phase gives
only a rather low superconducting transition temperature. This
puzzling point certainly warrants further investigations. Our data
here further illustrate that the superconductivity can be easily
induced by doping the Fe ions with many other transition metals
which are not restricted to the ones with 3d orbital electrons.

\subsection{Upper critical field}
\begin{figure}
\includegraphics[width=8cm]{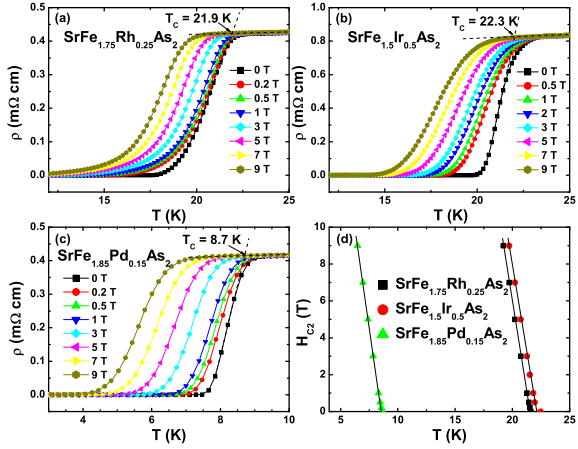}
\caption {(Color online) Temperature dependence of resistivity for
the samples (a) SrFe$_{1.75}$Rh$_{0.25}$As$_{2}$, (b)
SrFe$_{1.5}$Ir$_{0.5}$As$_{2}$, and (c)
SrFe$_{1.85}$Pd$_{0.15}$As$_{2}$ at different magnetic fields. The
dashed line indicates the extrapolated resistivity in the normal
state. One can see that the superconductivity seems to be robust
against the magnetic field and shifts slowly to the lower
temperatures. (d) The upper critical field determined using the
criterion of 90\%$\rho_n$.} \label{fig8}
\end{figure}

In Figs.~\ref{fig8}(a),~\ref{fig8}(b) and~\ref{fig8}(c) we present
the temperature dependence of resistivity for the samples
SrFe$_{1.75}$Rh$_{0.25}$As$_{2}$, SrFe$_{1.5}$Ir$_{0.5}$As$_{2}$ and
SrFe$_{1.85}$Pd$_{0.15}$As$_{2}$ under different magnetic fields.
Just as many other iron-pnictide superconductors, the
superconductivity is very robust against the magnetic field. We used
the criterion of $90\%\rho_n$ to determine the upper critical field
and show the data in Fig.~\ref{fig8}(d). The Slope of -dH$_{c2}$/dT
is 3.8 T/K for SrFe$_{1.75}$Rh$_{0.25}$As$_{2}$, 3.8 T/K for
SrFe$_{1.5}$Ir$_{0.5}$As$_{2}$, and 4.2 T/K for
SrFe$_{1.85}$Pd$_{0.15}$As$_{2}$, respectively. These values are
rather large which indicates rather high upper critical fields in
these systems. In order to determine the upper critical field in the
low temperature region, we adopted the Werthamer-Helfand-Hohenberg
(WHH) formula\cite{WHH} $H_{c2} = -0.69(dH_{c2}/dT)|_{T_c}T_c$. For
SrFe$_{1.75}$Rh$_{0.25}$As$_{2}$, by taking $(dH_{c2}/dT)|_{T_c}$ =
-3.8 T/K and $T_c$ = 21.9 K, and finally we have $H_{c2}(0)$ = 57.4
T. Similarly we get $H_{c2}(0)$ = 58 T for
SrFe$_{1.5}$Ir$_{0.5}$As$_{2}$ and 25.1 T for
SrFe$_{1.85}$Pd$_{0.15}$As$_{2}$. These $H_{c2}(0)$ values indicate
that the present 4d, 5d-transition metal doped samples have also
very large upper critical fields, as in K-doped\cite{WangZSPRB} and
Co-doped samples.\cite{Jo} Very recently the high upper critical
fields, as a common feature in the iron pnictide superconductors,
were interpreted as due to the strong disorder effect.\cite{HighHc2}

\subsection{Discussion}
The superconductivity mechanism in the FeAs-based superconductors
remains unclear yet. However, our present work and that with the Co
doping may give some hints on that. First of all, the three kind of
dopants (Co, Rh and Ir) reside in the same column in the periodic
table of elements. The relative atomic mass of these ions are quite
different: 58.9 for Co, 102.9 for Rh and 192.2 for Ir. Since these
atoms are doped into the FeAs-planes, they are certainly playing
important roles in governing the superconductivity. It is important
to note that doping the three different atoms into the system leads
to quite close maximum T$_c$s: 24 K for Co doping, 22 K for Rh
doping and 24 K for Ir doping. In the simple picture concerning the
electron-phonon coupling as the key mechanism for the pairing, the
Ir-doped sample should have the lowest T$_c$. We can even have a
brief estimate on T$_c$ based on the electron-phonon coupling
picture. For the Co-doped sample, the maximal T$_c$ appears at about
x = 0.16. In this case, we have a average mass for each Fe-site
(1.84*55.8+0.16*58.9)/(2Fe) = 56/Fe. Similarly in the Rh doped case,
the maximal T$_c$ appears at about x = 0.25, the average mass is
61.7/Fe. For Ir-doping, the maximal T$_c$ appears at about x = 0.43,
the average mass is 85.1/Fe. Using the relation of the isotope
effect $M^{\alpha}T_c = constant$ and taking $\alpha = 0.5$, we
would have T$_c$ (Co-doping):T$_c$ (Rh-doping):T$_c$ (Ir-doping) =
1: 0.95 :0.81. This is certainly far away from the experimental
values. Although the phonon spectrum as well as the electron band
structure will change with doping Co, Rh and Ir, above argument
should be qualitatively valid. In this sense, the experimental data
suggests that the three elements with very different mass lead to
negligible effect on the superconducting transition temperatures.
Actually our experiment naturally supports the picture that the
superconductivity is established by suppressing the AF order. The
related and widely perceived picture is that the pairing is through
the inter-pocket scattering of electrons via exchanging the AF spin
fluctuations.\cite{Mazin,Kuroki,WangF,WangZD} By doping electrons or
holes into the parent phase, the AF order will be destroyed
gradually. Instead, the short range AF order will provide a wide
spectrum of spin fluctuations which may play as the media for the
pairing between electrons. This picture can certainly give a
qualitative explanation to the occurrence of superconductivity in
the cases of doping Co, Rh and Ir.

\section{Conclusions}
Superconductivity has been observed in SrFe$_{2-x}$M$_x$As$_2$ (M =
Rh, Ir and Pd). For the three different dopants, Rh, Ir and Pd, it
was found that the normal state resistivity exhibits a roughly
linear behavior starting just above T$_c$ all the way up to 300 K at
the optimal doping point. This may reflect a novel scattering
mechanism in the normal state. The phase diagrams of
SrFe$_{2-x}$M$_x$As$_2$ (M = Rh, Ir and Pd) systems obtained are
quite similar to that by doping Co or Ni to the Fe sites. However,
the suppression to the AF order in doping Ir is much slower and the
superconductivity suddenly sets in at a high doping (x = 0.43).
Regarding the close maximal superconducting transition temperatures
in doping Co, Rh and Ir although they have very different masses, we
argue that the superconductivity is closely related to the
suppression of the AF order, rather than the electron-phonon
coupling. Through measuring the magnetic field induced broadening of
resistive transition curve we determined the upper critical field.
It is found that the superconductivity in all the doped samples is
rather robust against the magnetic field.

\section{Acknowledgements}
This work is supported by the Natural Science Foundation of China,
the Ministry of Science and Technology of China (973 project:
2006CB601000, 2006CB921802), the Knowledge Innovation Project of
Chinese Academy of Sciences (ITSNEM).

\end{document}